\input harvmac
\noblackbox
\def\npb#1#2#3{{\it Nucl.\ Phys.} {\bf B#1} (19#2) #3}
\def\plb#1#2#3{{\it Phys.\ Lett.} {\bf B#1} (19#2) #3}

\def\ijmp#1#2#3{{\it Int.\ J. Mod.\ Phys.} {\bf A#1} (19#2) #3}

\def\bar#1{\overline{#1}}

\def\frac#1#2{{#1 \over #2}}

\def\p{\partial}
\def\tilde#1{\widetilde{#1}}

                   \def\CG{{\cal G}}
                   
\def\CL{{\cal L}}

\def\R{{\bf R}}
\def\S{{\bf S}}
\def\X{{\bf X}}
\def\Z{{\bf Z}}
\Title{\vbox{\baselineskip12pt
\hbox{hep-th/9608019}
\hbox{PUPT-1637}
\hbox{}}}
{\vbox{\centerline{GLUINO CONDENSATION IN STRONGLY}
\bigskip
\centerline{COUPLED HETEROTIC STRING THEORY}}}
\smallskip
\centerline{Petr Ho\v rava\footnote{$^\ast$}{horava@puhep1.princeton.edu}}  
\medskip
\centerline{\it Joseph Henry Laboratories, Princeton University}
\centerline{\it Jadwin Hall, Princeton, NJ 08544, USA}
\baselineskip18pt
\bigskip\medskip
Strongly coupled heterotic $E_8\times E_8$ string theory, compactified to 
four dimensions on a large Calabi-Yau manifold $\X$,  may represent a viable 
candidate for the description of low-energy particle phenomenology.  
In this regime, heterotic string theory is adequately described by low-energy 
$M$-theory on $\R^4\times\S^1/\Z_2\times\X$, with the two $E_8$'s 
supported at the two boundaries of the world.  In this paper we study the 
effects of gluino condensation, as a mechanism for supersymmetry breaking in 
this $M$-theory regime.  We show that when a gluino condensate forms in 
$M$-theory, the conditions for unbroken supersymmetry can still be satisfied 
locally in the orbifold dimension $\S^1/\Z_2$.  Supersymmetry is then only 
broken by the global topology of the orbifold dimension, in a mechanism 
similar to the Casimir effect.  This mechanism leads to a natural hierarchy 
of scales, and elucidates some aspects of heterotic string theory that might 
be relevant to the stabilization of moduli and the smallness of the 
cosmological constant.  
\Date{July 1996}
\nref\chsw{P. Candelas, G. Horowitz, A. Strominger and E. Witten, ``Vacuum 
Configurations for Superstrings,'' \npb{258}{85}{46}.}
\nref\gsw{M.B. Green, J.H. Schwarz and E. Witten, ``Superstring Theory'' 
(Cambridge U. Press, 1987).}
\nref\dienes{for a recent review, see: K.R. Dienes, ``String Theory and the 
Path to Unification: A Review of Recent Developments,'' hep-th/9602045, to 
appear in {\it Phys.\ Rep.}}
\nref\peskin{M.E. Peskin, ``The Experimental Investigation of Supersymmetry 
Breaking,'' hep-ph/9604339.}
\nref\nillold{H.P. Nilles, ``Supersymmetry, Supergravity and Particle 
Physics,'' {\it Phys.\ Rep.} {\bf 110} (1984) 1.}
\nref\dineless{M. Dine, ``String Theory: Lessons for Low Energy Physics?,'' 
hep-th/9210047.}
\nref\dinatur{M. Dine, ``Problems of Naturalness: Some Lessons from String 
Theory,'' hep-th/9207045.}
\nref\modcosmo{T. Banks, M. Berkooz, S.H. Shenker, G. Moore and P.J. 
Steinhardt, ``Modular Cosmology,'' hp-th/9503114.}
\nref\banks{T. Banks, ``SUSY Breaking, Cosmology, Vacuum Selection and the 
Cosmological Constant in String Theory,'' hep-th/9601151.}
\nref\ewcosmo{E. Witten, ``Some Comments on String Dynamics,'' 
hep-th/9507121.}
\nref\weinberg{S. Weinberg, ``The Cosmological Constant Problem,'' 
{\it Rev.\ Mod.\ Phys.} {\bf 61} (1989) 1.}
\nref\disei{M. Dine and N. Seiberg, ``Is the Superstring Weakly Coupled?,'' 
\plb{162}{85}{299}; ``Is the Superstring Semiclassical?,'' in: Unified String 
Theories, eds: M.B.~Green and D.J.~Gross (World Scientific, 1986).}
\nref\bdstrong{T. Banks and M. Dine, ``Coping with Strongly Coupled String 
Theory,'' hep-th/9406132.}
\nref\dinestrong{M. Dine, ``Coming to Terms with Strongly Coupled Strings,'' 
hep-th/9508085.}
\nref\dinetruly{M. Dine and Y. Shirman, ``Truly Strong Coupling and Large 
Radius in String Theory,'' hep-th/9601175.}
\nref\hw{P. Ho\v rava and E. Witten, ``Heterotic and Type I String Dynamics 
from Eleven Dimensions,'' \npb{460}{96}{506}, hep-th/9510209.}
\nref\hweff{P. Ho\v rava and E. Witten, ``Eleven-Dimensional Supergravity 
on a Manifold with Boundary,'' hep-th/9603142; to appear in {\it Nucl.\ 
Phys.} {\bf B}.}
\nref\wstr{E. Witten, ``Strong Coupling Expansion of Calabi-Yau 
Compactification,'' hep-th/9602070.}
\nref\polchinski{J. Polchinski, ''String Duality -- A Colloquium,'' 
hep-th/9607050.}
\nref\banksdine{T. Banks and M. Dine, ``Couplings and Scales in Strongly 
Coupled Heterotic String Theory,'' hep-th/9605136.}
\nref\kaplun{E. Caceres, V.S. Kaplunovsky and I.M. Mandelberg, 
``Large-Volume String Compactifications, Revisited,'' hep-th/9606036.}
\nref\ewsuper{E. Witten, ``Non-Perturbative Superpotentials in String 
Theory,'' hep-th/9604030.}
\nref\nillsugra{H.P. Nilles, ``Dynamically Broken Supergravity and the 
Hierarchy Problem,'' \plb{115}{82}{193}.}
\nref\drsw{M. Dine, R. Rohm, N. Seiberg and E. Witten, ``Gluino Condensation 
in Superstring Models,'' \plb{156}{85}{55}.}
\nref\dinill{J.P. Derendinger, L.E. Ib\'a\~nez and H.P. Nilles, ``On the 
Low Energy $d=4$, $N=1$ Supergravity Theory Extracted from the $d=10$, 
$N=1$ Superstring,'' \plb{155}{85}{65}.}
\nref\glurevnill{H.P. Nilles, ``Gaugino Condensation and Supersymmetry 
Breakdown,'' \ijmp{5}{90}{4199}.}
\nref\condreviews{F. Quevedo, ``Gaugino Condensation, Duality and 
Supersymmetry Breaking,'' hep-th/9511131;\hfill\break
F. Quevedo, ``Lectures on Superstring Phenomenology,'' 
hep-th/9603074;\hfill\break
H.P. Nilles, ``Dynamical Gauge Coupling Constants,'' 
hep-ph/960124.}
\nref\krasnikov{N.V. Krasnikov, ``On Supersymmetry Breaking in Superstring 
Theories,'' \plb{193}{87}{37}.}
\nref\dixon{L.J. Dixon, ``Supersymmetry Breaking in String Theory,'' talk at 
the APS meeting in Houston, SLAC preprint (April 1990).}
\nref\njunk{Z. Lalak, A. Niemeyer and H.P. Nilles, ``Gaugino Condensation, 
S-Duality and Supersymmetry Breaking in Supergravity Models,'' hep-th/9503170;
\hfill\break
A. Niemeyer and H.P. Nilles, ``Gaugino Condensation and the Vacuum 
Expextation Value of the Dilaton,'' hep-th/9508173.}
\nref\dqjunk{C.P. Burgess, J.-P. Derendinger, F. Quevedo and M. Quir\'os, 
``On Gaugino Condensation with Field-Dependent Gauge Couplings,'' 
hep-th/9505171.}
\nref\cremmer{E. Cremmer, S. Ferrara, C. Kounnas and D.V. Nanopoulos, 
``Naturally Vanishing Cosmological Constant in $N=1$ Supergravity,'' 
\plb{133}{83}{61}.}
\nref\nanop{J.Ellis, C. Kounnas and D.V. Nanopoulos, ``Phenomenological 
$SU(1,1)$ Supergravity,'' \npb{241}{84}{406};\hfill\break
J.Ellis, A.B. Lahanas, D.V. Nanopoulos and K. Tamvakis, ``No-Scale 
Supersymmetric Standard Model,'' \plb{134}{84}{429};\hfill\break
J.Ellis, C. Kounnas and D.V. Nanopoulos, ``No-Scale Supersymmetric GUTs,'' 
\npb{247}{84}{373};\hfill\break
for a review, see:  A.B. Lahanas and D.V. Nanopoulos, ``The Road to 
No-Scale Supergravity,'' {\it Phys.\ Rep.} {\bf 145} (1987) 1.}
\newsec{Introduction: Phenomenology of $M$-Theory}

Despite its remarkable phenomenological promise \refs{\chsw,\gsw,\dienes}, 
string theory still 
leaves unanswered many pressing questions about its contact with the 
low-energy world.  Among the issues that we would certainly want to 
understand better in a unified theory are the mechanism of supersymmetry 
breaking with a large hierarchy of scales, the stabilization of moduli, and 
the smallness of the cosmological constant (for reviews and references, 
see \refs{\peskin-\weinberg}).  

Our present understanding of this subject indicates that string theory might 
be able to identify the right degrees of freedom in which phenomenology can 
be naturally understood.  There are, however, equally strong indications that 
in the regime directly relevant to phenomenology, the natural degrees of 
freedom of perturbative string theory are strongly coupled 
\refs{\disei-\dinetruly}.  Recently, 
we have witnessed a revolution that is rapidly changing our understanding of 
string theory in the strongly coupled regime, leading in many cases to a dual 
decription of the physics in this regime in terms of more natural, weakly 
coupled degrees of freedom.  One can wonder whether these dual descriptions 
might lead to variables more appropriate for the description of low-energy 
phenomenology.  

In the recent studies of string dualities, at least one such new paradigm 
may have already appeared \refs{\hw-\wstr}.  Compactification of the 
strongly coupled 
heterotic $E_8\times E_8$ string theory on a Calabi-Yau manifold $\X$ is most 
naturally described by eleven-dimensional $M$-theory, compactified to $\R^4$ 
on a 
manifold with one extra dimension, $\X\times\S^1/\Z_2$ \hw .  This extra 
dimension 
-- invisible at weak heterotic coupling -- is an orbifold dimension, and 
the total space-time manifold has a boundary with two 
components.  At low energies, the effective description of $M$-theory is in 
terms of eleven-dimensional 
supergravity, coupled to one $E_8$ Yang-Mills supermultiplet at each boundary 
of the world \hweff .  

This picture 
gives an interesting new twist to the old Kaluza-Klein idea.  For a 
low-energy observer, the world first looks four-dimensional.  After crossing 
a certain threshold, the world becomes effectively five-dimensional, but the 
matter sector containing the standard model still lives at a four-dimensional 
boundary.  The bulk of the 
five-dimensional space-time supports only gravity (as well as 
other fields coming 
from the eleven-dimensional supergravity multiplet).  At the other end of 
the world, another gauge sector -- the other $E_8$ of the heterotic 
string theory -- is hidden, and communicates with the matter of the standard 
model only gravitationally.  Finally, at even higher energies, the observer 
reaches the compactification scale and sees the additional six dimensions 
compactified on the Calabi-Yau manifold, and the world becomes 
eleven-dimensional.  

This newly understood regime of heterotic string theory seems very attractive 
phenomenologically.  In \wstr , Witten used this regime to analyze the 
strongly coupled heterotic string compactified on a large 
Calabi-Yau manifold, with the four-dimensional grand-unified 
coupling $\alpha_{\rm GUT}$ acceptably small.  
A detailed analysis reveals that for such compactifications -- unlike in the 
weakly coupled heterotic string -- the strengths of all interactions including 
gravitational can be naturally unified at the unification scale.  In other 
words, the unacceptable prediction of the size of the Newton constant $G_N$ -- 
as made generically by the weakly coupled heterotic string theory -- is 
alleviated at strong coupling, in the $M$-theory regime.%
\foot{{}For a clear non-technical 
exposition of this result, see Section~4.3 and Figure~6 of \polchinski.}
More recently, other interesting phenomenological implications of this 
scenario have been studied in some detail in \banksdine\ (see also \kaplun ).  

\subsec{Gluino Condensate, Supersymmetry Breaking, and the Cosmological 
Constant}

The unification of couplings -- essentially, due to the presence of the extra 
dimension of the type discussed in \hw\ -- can be condsidered one of the first 
phenomenological successes of $M$-theory. This makes one wonder whether 
$M$-theory has anything to say 
about the mysteries of hierarchical supersymmetry breaking and the smallness 
of the cosmological constant.  

There are three known mechanisms that can trigger supersymmetry breaking 
in string theory:  world-sheet instantons, space-time instantons, and strong 
infrared dynamics.  Superpotentials generated by instantons have been 
recently studied in \ewsuper , where three-dimensional compactifications of 
$F$-theory were used to gain information about four-dimensional physics.  

Gluino condensation in the hidden sector is a representant of the third 
class.  It has been extensively studied as a mechanism of supersymmetry 
breaking, ever since the pioneering papers in supergravity \nillsugra\ and in 
string theory \refs{\drsw,\dinill}; for reviews and references on this 
subject, see \refs{\glurevnill,\condreviews}.%
\foot{The simplest version of gluino condensation in weakly coupled heterotic 
string theory  does not successfully explain the hierarchy of scales and the 
stabilization of moduli.  Several suggestions how to alleviate 
these problems without abandoning the regime of weak string coupling have 
been made \dienes , among them the racetrack models with multiple gluino 
condensates \refs{\krasnikov,\dixon}, or modified gauge kinetic functions 
\refs{\njunk,\dqjunk}.}

In addition to providing a natural mechanism of supersymmetry breaking, gluino 
condensation could also be relevant to the cosmological constant problem and 
the stabilization of moduli in string theory.  At a very 
early stage of the studies of gluino condensates in weakly coupled heterotic 
string theory, it has been noticed \refs{\drsw,\dinill} 
that certain terms in the Lagrangian of 
the ten-dimensional heterotic supergravity conspire in a very particular way, 
leading to a potential of a very special, ``no-scale'' type, first considered 
in \refs{\cremmer,\nanop}.  
Potentials of the no-scale type have been argued to break supersymmetry 
while keeping the cosmological constant naturally zero without fine tuning.  
The main problem with this mechanism seems to be the apparent absence of a 
satisfactory symmetry principle that could explain and protect this 
particular form of the potential and lead to supersymmetry breaking with 
zero cosmological constant in the presence of quantum effects.  

In this paper, we will study gluino condensation in the strongly coupled, 
$M$-theory regime of the heterotic string theory, to the lowest non-trivial 
order in a long-wavelength expansion.  We will assume that a 
gluino condensate develops in the $E_8$ sector hidden at the other end of 
the world, and will study its consequences for supersymmetry breaking.%
\foot{The formation of a gluino condensate is exactly what one expects on the 
basis of a simple physical argument.  {}For the compactifications studied in 
\wstr\ and in the present paper, strong coupling in the hidden $E_8$ develops 
exactly when the other couplings attain phenomenologically interesting 
values.  With strong gauge coupling in the hidden sector, we can expect that 
a gluino condensate is dynamically generated.  This aspect of the strong 
gauge dynamics should be stable under the effects caused by the coupling to 
gravity.}
Our analysis will reveal some unexpected properties of gluino condensation in 
the 
heterotic string theory at strong coupling.  Since the rest of the paper will 
be somewhat technical, we summarize our results here: 

In the low-energy Lagrangian of $M$-theory, we find a ``conspiration of 
terms'' similar to the one observed in the low-energy Lagrangian of the weakly 
coupled heterotic string theory.  Incidentally, this explains some rather 
singular terms encountered in \hweff\ in the construction of the low-energy 
effective Lagrangian of $M$-theory on a manifold with boundary.  It also 
suggests that when a gluino condensate develops at the boundary, the field 
strength $G$ of the three-form $C$ from the eleven-dimensional supergravity 
multiplet develops a compensating vacuum expectation value supported at 
the boundary.  We also encounter first indications that the eleven-dimensional 
variables of $M$-theory might be more appropriate for the description of 
supersymmetry breaking by the gluino condensate -- the role of the would-be 
goldstino is played by the normal component $\psi_{11}$ of the 
eleven-dimensional gravitino.  

Even in the presence of the gluino condensate, we will 
still be able to solve the unbroken supersymmetry conditions in any given 
coordinate system.  This phenomenon might come as a 
surprise, and is intimately related to the existence of the extra dimension
in $M$-theory.  

The solution of unbroken supersymmetry conditions exists locally, but not 
globally in the extra dimension.  When we try to extend the local solution 
globally over $\S^1/\Z_2\times\X$, we encounter a topological obstruction 
(essentially, the total cohomology class of the gluino condensate).  
Therefore, supersymmetry is broken by the global topology of 
the extra, orbifold dimension, in a process similar to the Casimir effect.  

The fact that the unbroken supersymmetry conditions can be satisfied in any 
coordinate system on 
$\R^4\times\S^1/\Z_2\times\X$ leads to an intriguing refinement of the 
phenomenology of supersymmetry 
breaking in these models.  We have argued that in the $M$-theory scenario, 
observers 
at intermediate energies will see the world as five-dimensional.  At length 
scales larger than the Calabi-Yau compactification radius but still much 
smaller than the radius of the fifth dimension, these observers should see 
unbroken supersymmetry, even if they are directly at the other end of the 
world where the gluino condensate has formed.  As the resolution is 
diminished, the supersymmetry breaking effects -- caused by the compactness 
of the fifth dimension -- become visible, and at even larger length scales, 
the world will be effectively four-dimensional and supersymmetry will be 
broken.  However, this breakdown should be rather mild, since it is only 
caused by effects sensitive to the global topology of the fifth dimension.  

This mechanism of supersymmetry breaking generates a natural 
hierarchy of scales.  In the phenomenologically interesting regime, the 
radius of the fifth dimension can be expected \banksdine\  
to be at least an order of magnitude larger than the eleven-dimensional 
Planck length.  The mass of 
the five-dimensional gravitino is only induced quantum mechanically, by 
loop effects sensitive to the size of the fifth dimension, and is therefore 
suppressed by a power of the inverse radius of the fifth dimension.  

This hidden eleven-dimensional supersymmetry -- broken only by the 
global topology of the orbifold dimension -- explains the 
``conspiracy'' that leads in the weakly coupled heterotic string theory 
to the no-scale 
potential with supersymmetry breaking and zero cosmological constant 
at tree level.  

\newsec{Gluino Condensation in Heterotic String Theory and $M$-Theory}

In this paper we will study the heterotic $E_8\times E_8$ string theory 
compactified to $\R^4$ on a Calabi-Yau three-fold $\X$.  In the strong 
coupling limit, the low-energy description of this theory is in terms 
of eleven-dimensional $M$-theory compactified on 
$\R^4\times\S^1/\Z_2\times\X$, with the two $E_8$ gauge groups supported at 
the two space-time boundaries in the orbifold dimension $\S^1/\Z_2$.  We will 
deal with various supergravities that describe the low-energy physics of such 
compactifications.  

Our ten-dimensional and eleven-dimensional conventions are as in \hw .  The 
space-time signature is $-+\ldots+$.  Eleven-dimensional vector indices will 
be written as $I,J,K,\ldots$.   The eleven-dimensional gamma matrices 
are $32\times 32$ real matrices satisfying $\{\Gamma_I,\Gamma_J\}=2g_{IJ}$, 
with $g_{IJ}=\eta_{mn}e_I{}^me_J{}^n$ the eleven-dimensional metric.  Each 
of the two boundary components of the eleven-dimensional manifold supports 
one $E_8$ Yang-Mills supermultiplet.  One of the $E_8$'s will be broken by 
the spin connection embedding to a grand-unified $E_6$ group, while the 
other $E_8$ will be strongly coupled and hidden at the other end of the 
world.  The adjoint index of this hidden $E_8$ will be denoted by 
$a,b,\ldots$.  

On $\R^4\times\S^1/\Z_2\times\X$, we will use four-dimensional vector indices 
$\mu,\nu,\ldots$ that parametrize the flat Minkowski space $\R^4$, and vector 
indices $i,j,k,\ldots$ and their compex conjugates $\bar\imath,\bar\jmath,
\bar k,\ldots$ that correspond to a complex coordinate system on the 
Calabi-Yau three-fold $\X$.  The ten-dimensional vector indices that 
parametrize $\R^4\times\X$ will be written as $A,B,C,\ldots$.  Our other 
conventions on $\X\times\S^1/\Z_2$ are as in \wstr .  

\subsec{Gluino Condensation and the Potential at Weak Coupling}

First we recall some aspects of the gluino condensation in the hidden sector 
of the weakly coupled heterotic string theory that will be relevant for our 
purposes.  

Consider, as in \drsw , the weakly coupled heterotic $E_8\times E_8$ string 
theory compactified on a Calabi-Yau three-fold $\X$.   On any given $\X$, we 
have a covariantly constant holomorphic three-form $\epsilon_{ijk}$  (and 
its anti-holomorphic complex conjugate $\bar\epsilon_{\bar\imath
\bar\jmath\bar k}$).  In ten dimensions, $\bar\chi^a\Gamma_{ABC}\chi^a$ is 
the only gluino bilinear that is not identically zero by fermi statistics 
and chirality.  If this bilinear develops a non-zero vacuum expectation value 
proportional to the covariantly constant holomorphic three-form on $\X$, 
\eqn\eeglucond{\left\langle\bar\chi^a\Gamma_{ijk}\chi^a\right\rangle=
c\Lambda_{E_8}^3\epsilon_{ijk}}
(and similarly for the complex conjugate), the four-dimensional observer will 
interpret this expectation value as a non-zero gluino condensate, $\left
\langle\bar\chi^a\chi^a\right\rangle$ (and $\left\langle\bar\chi^a\gamma_5
\chi^a\right\rangle$).  In \eeglucond , $\Lambda_{E_8}$ 
is the characteristic scale of 
the hidden gauge sector, at which the gauge coupling becomes strong, and $c$ 
is a (complex) number of order one.  

In the process of analyzing the physics of the gluino condensate in weakly
coupled heterotic string theory, it has been noticed \drsw\ (see also \gsw , 
Vol.\ 2, p.\ 326) that the Lagrangian of
ten-dimensional heterotic supergravity exhibits a special feature that 
could lead -- at least at tree level -- to supersymmetry breaking with zero 
cosmological constant without fine tuning.  The argument is roughly as 
follows.  The Lagrangian contains a gluino self-interaction term which is 
quartic in $\chi^a$; it also contains an interaction between the gluino 
bilinear $\bar\chi^a\Gamma_{ABC}\chi^a$ and the three-form field strength 
$H_{ABC}$.  Together with the kinetic term $H^2$, these 
terms conspire in such a way that they can be assembled into a perfect 
square, 
\eqn\eepersquare{-\frac{3\kappa_{10}^2}{4\lambda_{10}^4}\int_{M^{10}}d^{10}x
\sqrt g\frac{1}{\phi^{3/2}}
\left(H_{ABC}-\lambda_{10}^2\sqrt 2\phi^{3/4}(\bar\chi^a
\Gamma_{ABC}\chi^a)\right)^2.}  
(We have used the normalizations of \drsw ; $\phi$ is the ten-dimensional 
dilaton, while $\kappa_{10}$ and $\lambda_{10}$ denote the ten-dimensional 
gravitational and gauge coupling, respectively.)  Consider now the situation 
in which a gluino condensate has formed, proportional to the covariantly 
constant three-form on $\X$ as in \eeglucond .  If we assume 
that the three-form field strength $H_{ABC}$ develops a compensating vacuum 
expectation value, 
\eqn\eehcond{H_{ijk}=c\Lambda_{E_8}^3\lambda_{10}^2\sqrt 2\phi^{3/4}
\epsilon_{ijk},}
such that the perfect square term \eepersquare\ in the potential vanishes, the 
cosmological constant will be zero at tree level.  At the same time, one can 
show that supersymmetry is broken by the condensates \eeglucond\ and 
\eehcond .  

The easiest way to see the supersymmetry breaking in the presence of the 
condensates is to look at the relevant part of the supersymmetry variation 
of the fermions.  There are two relevant fermions in the 
theory -- the ten-dimensional gravitino $\Psi_A$, and the dilatino 
$\lambda$.  Schematically, the relevant parts of their supersymmetry 
variations are given by
\eqn\eetenvar{\eqalign{\delta\Psi_A&=\frac{1}{\kappa_{10}}D_A\eta
+\frac{\sqrt 2}{32}\left(\frac{\kappa_{10}}{\lambda_{10}^2}\right)
\frac{1}{\phi^{3/4}}H_{BCD}\left(\Gamma_A{}^{BCD}-9\delta_A^B\Gamma^{CD}\right)
\eta\cr
&\qquad{}-\frac{1}{256}\kappa_{10}\,\left(\bar\chi^a\Gamma_{BCD}\chi^a
\right)\left(\Gamma_A{}^{BCD}-5\delta_A^B\Gamma^{CD}\right)\eta+\ldots,\cr
\delta\lambda&=\ldots
+\frac{1}{8}\left(\frac{\kappa_{10}}{\lambda_{10}^2}\right)
\frac{1}{\phi^{3/4}} H_{ABC}\Gamma^{ABC}\eta
+\frac{\sqrt 2}{384}\kappa_{10}\,\left(\bar\chi^a\Gamma_{ABC}\chi^a\right)
\Gamma^{ABC}\eta+\ldots.\cr}}
(Here the $\ldots$ correspond to terms that are either proportional to the 
gravitino and dilatino, or contain the space-time derivative of the 
dilaton.)  
We can see from \eetenvar\ that in the presence of the condensates 
\eeglucond\ and \eehcond , the unbroken supersymmetry conditions 
$\delta\Psi_A=0$ and $\delta\lambda=0$ cannot be satisfied.  A particular 
linear combination of $\Psi_A$ and $\lambda$ behaves as a Goldstone 
fermion and gives a non-zero tree-level mass to the gravitino, and 
supersymmetry is broken in this approximation.  

In four dimensions, the perfect square structure \eepersquare\ of the 
heterotic supergravity Lagrangian leads to the superpotential and K\"ahler 
potential of the very special, no-scale type \refs{\cremmer,\nanop}.  
Superpotentials and 
K\"ahler potentials of the no-scale type were proposed \cremmer\ 
in earlier attempts to link supersymmetry 
breaking with the solution of the cosmological constant problem (see also the 
discussion in \weinberg).  One of the main drawbacks of this approach so far 
has been the apparent lack of a symmetry principle that could explain and 
protect this particular form of the potential.%
\foot{Recently, some attempts have been made \njunk\ to substantiate 
the no-scale potentials using $S$-duality.}

\subsec{Strong Coupling and $M$-Theory}

At strong string coupling and large radius of the Calabi-Yau manifold, the 
compactification is effectively described by low-energy $M$-theory on 
$\R^4\times\S^1/\Z_2\times\X$.  The effective Lagrangian for this theory has 
been constucted in \hweff .  It contains the eleven-dimensional supergravity 
multiplet $e_I{}^m,\psi_J$ and $C_{IJK}$ in the bulk, coupled to one $E_8$ 
Yang-Mills supermultiplet $A_B^a,\chi^a$ at each of the two ten-dimensional 
boundaries.  

To order $\kappa^{2/3}$, the Lagrangian is given by
\eqn\mthlag{\eqalign{\CL&=\frac{1}{\kappa^2}\int_{M^{11}}d^{11}x\sqrt g
\left(-\frac{1}{2}R-\frac{1}{2}\bar\psi_I\Gamma^{IJK}D_J\left(\frac{\Omega+
\hat\Omega}{2}\right)\psi_K-\frac{1}{48}
G_{IJKL}G^{IJKL}\right.\cr
&\qquad\qquad{}-\frac{\sqrt 2}{384}\left(\bar\psi_I\Gamma^{IJKLMN}\psi_N
+12\bar\psi^J\Gamma^{KL}\psi^M\right)\left(G_{JKLM}+\hat G_{JKLM}\right)\cr
&\qquad\qquad\qquad\left.{}-\frac{\sqrt 2}{3456}\epsilon^{I_1I_2\ldots I_{11}}
C_{I_1I_2I_3}G_{I_4\ldots I_7}G_{I_8\ldots I_{11}}\right)\cr
&{}+\frac{1}{2\pi(4\pi\kappa^2)^{2/3}}\int_{M^{10}}d^{10}x\sqrt g\left(
-\frac{1}{4}F^a_{AB}F^{a\,AB}-\frac{1}{2}\bar\chi^a\Gamma^AD_A(\hat\Omega)
\chi^a\right.\cr
&\left.\qquad{}-\frac{1}{8}\bar\psi_A\Gamma^{BC}\Gamma^A\left(F^a_{BC}+
\hat F^a_{BC}\right)\chi^a+\frac{\sqrt 2}{48}\left(\bar\chi^a\Gamma^{ABC}
\chi^a\right)\hat G_{ABC\,11}\right).\cr}}
(Explicit expressions for the supercovariant objects $\hat\Omega$, $\hat 
F^a_{AB}$ and $\hat G_{IJKL}$ can be found in \hweff .)  The fields of the 
bulk supergravity multiplet satisfy natural orbifold boundary conditions, 
discussed in detail in \hweff .  It was also shown in \hweff\ that the
four-form field strength $G_{IJKL}$ satisfies a modified Bianchi identity, 
\eqn\eemodbione{dG_{11\,ABC}=-\frac{3\sqrt 2}{2\pi}\left(\frac{\kappa}{4\pi}
\right)^{2/3}\delta(x^{11})\left(\tr\;F_{[AB}F_{CD]}-\frac{1}{2}R_{[AB}R_{CD]}
\right),}
which will be important later in the paper.  

The effective Lagrangian \mthlag\ is invariant under local supersymmetry, 
whose parameter $\eta$ satisfies 
the orbifold condition $\eta(-x^{11})=\Gamma_{11}\eta(x^{11})$.  For the 
purposes of this paper, we will only need the rules for the supersymmetry 
transformations of the fermions; the relevant supersymmetry transformations 
are 
\eqn\eemthsusy{\eqalign{\delta\psi_A&=D_A\eta+\frac{\sqrt 2}{288}G_{IJKL}
\left(\Gamma_A{}^{IJKL}-8\delta_A^I\Gamma^{JKL}\right)\eta\cr
&\qquad{}-\frac{1}{576\pi}\left(\frac{\kappa}{4\pi}\right)^{2/3}\delta(x^{11})
\left(\bar\chi^a\Gamma_{BCD}\chi^a\right)\left(\Gamma_A{}^{BCD}-6\delta_A^B
\Gamma^{CD}\right)\eta+\ldots,\cr
\delta\psi_{11}&=D_{11}\eta+\frac{\sqrt 2}{288}G_{IJKL}\left(
\Gamma_{11}{}^{IJKL}-8\delta_{11}^I\Gamma^{JKL}\right)\eta
\cr
&\qquad{}+\frac{1}{576\pi}\left(\frac{\kappa}{4\pi}\right)^{2/3}
\delta(x^{11})\left(\bar\chi^a\Gamma_{ABC}\chi^a\right)\Gamma^{ABC}\eta+
\ldots,\cr
\delta\chi^a&=-\frac{1}{4}F^a_{AB}\Gamma^{AB}\eta+\ldots.\cr}}
The $\ldots$ denote terms of order $\kappa^{4/3}$, as well as known terms 
of order $\kappa^{2/3}$ bilinear in the gravitinos that we will not need.  
  
As we recalled in the previous subsection, the effective supergravity 
Lagrangian of the weakly coupled ten-dimensional heterotic string theory 
describes 
the interaction between the gluino bilinears $\bar\chi^a\Gamma_{ABC}
\chi^a$ and the three-form field strength $H_{ABC}$, by the 
perfect square term \eepersquare\ 
-- leading to the no-scale potential and the 
corresponding mechanism of supersymmetry breaking.  
At first, one would not expect such a perfect square structure to also appear 
in the effective Lagrangian of $M$-theory.  Indeed, the gluinos of $M$-theory 
live at the space-time boundary and can only contribute to the 
Lagrangian through surface terms.  On the other hand, the three-form 
$C_{IJK}$ -- whose field strength four-form $G_{IJKL}$ is the $M$-theory 
counterpart of the heterotic field strength $H_{ABC}$ -- belongs to the 
supergravity multiplet, and its kinetic term is supported by the 
bulk of the eleven-dimensional manifold.  We have indeed seen that the 
effective Lagrangian \mthlag\ contains the corresponding terms, 
\eqn\eeterms{-\frac{1}{12\kappa^2}\int_{M^{11}}d^{11}x\sqrt g\;
G_{ABC\,11}^2+\frac{\sqrt 2}{24(4\pi)^{5/3}\kappa^{4/3}}\int_{M^{10}}d^{10}x
\sqrt g\,G_{11\,ABC}\left(\bar\chi^a\Gamma^{ABC}\chi^a\right).}

Nevertheless, it is intriguing to notice that in fact, the 
perfect square structure of the interaction between the gluinos and the 
bosonic field strength persists also in $M$-theory.  In the construction of 
the Lagrangian \hweff\ an unusual boundary interaction term was encountered.   
This term appears at relative order $\kappa^{4/3}$, is quartic in the 
gluinos, and most importantly, is proportional to the boundary delta function 
$\delta(x^{11})$ evaluated at zero:  
\eqn\eedelzero{-\frac{\delta(0)}{96(4\pi)^{10/3}\kappa^{2/3}}
\int_{M^{10}}d^{10}x\sqrt g\left(\bar\chi^a\Gamma_{ABC}\chi^a\right)^2.}
In \hweff , the presence of this term in the effective Lagrangian has been 
inferred from the requirement of local supersymmetry.  That argument was 
rather formal and involved cancellations of infinities.  Still, it is 
interesting that this term turned out \hweff\ with precisely the right 
coefficient so that it can be combined with the two terms in \eeterms\ into 
a perfect square: 
\eqn\eemthpersq{\eqalign{-\frac{1}{12\kappa^2}&\int_{M^{11}}d^{11}x\sqrt g\;
G_{ABC\,11}^2+\frac{\sqrt 2}{24(4\pi)^{5/3}\kappa^{4/3}}\int_{M^{10}}d^{10}x
\sqrt gG_{11\,ABC}\left(\bar\chi^a\Gamma^{ABC}\chi^a\right)\cr
&\qquad\qquad\qquad{}-\frac{\delta(0)}{96(4\pi)^{10/3}
\kappa^{2/3}}
\int_{M^{10}}d^{10}x\sqrt g\left(\bar\chi^a\Gamma_{ABC}\chi^a\right)^2\cr
&{}=-\frac{1}{12\kappa^2}\int_{M^{11}}d^{11}x\sqrt g\left(
G_{ABC\,11}-\frac{\sqrt 2}{16\pi}\left(\frac{\kappa}{4\pi}\right)^{2/3}
\delta(x^{11})\,\bar\chi^a\Gamma_{ABC}\chi^a\right)^2.\cr}}

Of course, we can turn this argument around, and claim that the perfect 
square structure of the Lagrangian provides a rationale for the existence 
of the rather singular term \eedelzero\ in the effective Lagrangian of 
\hweff .  This statement can be given the following more precise meaning.  
Inspired by the perfect square structure of the Lagrangian as found in 
\eemthpersq , we can reassemble terms in the Lagrangian and redefine the 
fields, so that the Lagrangian and the supersymmetry transformations no 
longer contains any explicit terms 
proportional to infinite coefficients such as $\delta(0)$.  In what follows, 
we will shift the field strength four-form $G_{IJKL}$ by a term supported 
at the boundary and bilinear in the gluinos, and define a modified field 
strength $\tilde G_{IJKL}$ by 
\eqn\eemodifg{\eqalign{\tilde G_{ABC\,11}&=G_{ABC\,11}-\frac{\sqrt 2}{16\pi}
\left(\frac{\kappa}{4\pi}\right)^{2/3}\delta(x^{11})\bar\chi^a\Gamma_{ABC}
\chi^a,\cr
\tilde G_{ABCD}&=G_{ABCD}.\cr}}
This set of redefined fields is probably better suited for the description  
of the physics at the relevant scales, since it makes the effective 
Lagrangian free of formal infinities, to the order to which the low-energy 
field theory was claimed to make sense in \hweff .  

In the next section we will be interested in configurations on $\R^4\times
\S^1/\Z_2\times\X$ that preserve four-dimensional Poincar\'e invariance.  In 
those cases, all components $\tilde G_{\mu JKL}$ -- with $\mu$ the vector 
index on $\R^4$ -- will vanish.  The equations of motion for the non-zero 
components of the 
modified field strength on $\R^4\times\X\times\S^1/\Z_2$ are then 
\eqn\eemodifeom{D_I\tilde G^{IJKL}=0,}
i.e.\ they formally coincide with the equations of motion for the unmodified 
field strength $G_{IJKL}$ in the absence of the gluino condensate.  Of course, 
this fact depends crucially on the perfect square structure of the 
Lagrangian.  

The field strength $G_{IJKL}$ of the three-form $C$ has to satisfy the Bianchi 
identity \eemodbione .  In the transformed variables, the Bianchi identity 
becomes
\eqn\eemodifbi{\eqalign{d\tilde G_{11\,ABCD}&=-\frac{3\sqrt 2}{2\pi}\left(
\frac{\kappa}{4\pi}\right)^{2/3}\delta(x^{11})\left(\tr\;F_{[AB}F_{CD]}-
\frac{1}{2}R_{[AB}R_{CD]}\right)\cr
&\qquad{}+\frac{\sqrt 2}{4\pi}
\left(\frac{\kappa}{4\pi}\right)^{2/3}\delta(x^{11})\;
\p_{\,[A}\left(\bar\chi^a\Gamma_{BCD]}\chi^a\right).\cr}}
For a covariantly constant gluino condensate -- such as the one in 
\eeglucond , proportional to the covariantly constant holomorphic three-form 
$\epsilon_{ijk}$ on $\X$ -- the last term in \eemodifbi\ vanishes 
identically.  The Bianchi identity then formally coincides with the Bianchi 
identity for the unmodified field strength $G_{IJKL}$ in the absence of the 
gluino condensate. 

\newsec{Gluino Condensate and Supersymmetry in $M$-Theory}

Now we would like to solve the equations of motion in the 
presence of the gluino condensate.  Our strategy will be as follows.  First 
we find a solution of the equations of motion and the Bianchi identity for 
the four-form $\tilde G_{IJKL}$.  Then we will try to solve the conditions for 
unbroken supersymmetry, a priori expecting an obstruction that should prevent 
us from finding unbroken supersymmetry in the presence of a gluino 
condensate.  It will come as a surprise that -- because of the presence of 
the extra orbifold dimension of $M$-theory --  the unbroken supersymmetry 
conditions {\it can\/} actually be satisfied, locally in the extra 
dimension.  The expected obstruction will only be topological in nature, and 
will prevent us from extending the local solution globally over the extra 
dimension.  

As a first step, we have to solve the equations of motion and the Bianchi 
identity for the four-form field strength, which in the presence of a 
covariantly constant gluino condensate on $\X$ are
\eqn\eebieom{\eqalign{D^I\tilde G_{IJKL}&=0,\cr
d\tilde G_{ABCD\,11}&=-\frac{3\sqrt 2}{2\pi}\left(\frac{\kappa}{4\pi}
\right)^{2/3}\delta(x^{11})\left(\tr\;F_{[AB}F_{CD]}-\frac{1}{2}R_{[AB}R_{CD]}
\right)\cr
&\quad{}-\frac{3\sqrt 2}{2\pi}\left(\frac{\kappa}{4\pi}
\right)^{2/3}\delta(x^{11}-R_{11})\left(\tr\;F_{[AB}F_{CD]}-\frac{1}{2}
R_{[AB}R_{CD]}\right).\cr}}
(Here we have explicitly included -- unlike in our previous discussion -- 
the contribution from the other boundary, located at $x^{11}=R_{11}$.) 

The Bianchi identity in \eebieom\ cannot be satisfied unless the total 
cohomology class of its right hand side vanishes, 
\eqn\eecohom{\sum[F\wedge F]-[R\wedge R]=0.}
In the compactifications most directly relevant to phenomenology, this 
condition is satisfied by embedding the spin connection into one of the gauge 
groups, which is then broken from $E_8$ to $E_6$.  This embedding makes 
$\tr\,F\wedge F-R\wedge R$ vanish pointwise in the Calabi-Yau manifold, but 
does not make the right hand side of the Bianchi identity in \eebieom\ zero 
pointwise.  As argued in \wstr , this generates a gradient for the four-form 
field strength, which is therefore generically non-zero in this particular 
class of $M$-theory compactifications.%
\foot{This fact could be relevant to the stabilization of moduli in 
$M$-theory.}
Since the source of $d\tilde G$ is 
of order $\kappa^{2/3}$ in the long-wavelength expansion, $\tilde G_{IJKL}$ 
will also be of order $\kappa^{2/3}$.  

We have seen in the previous section that the equations \eebieom\ for 
$\tilde G_{IJKL}$ in the presence of a covariantly constant condensate 
coincide with the equations for the unmodified four-form $G_{IJKL}$ in the 
absence of the condensate.  These equations have been solved -- to the same 
order in $\kappa^{2/3}$ that we are interested in -- by Witten in \wstr . 
To solve \eebieom , we can take any solution $\CG_{IJKL}$ 
from \wstr , and set 
\eqn\eesolut{\tilde G_{IJKL}=\CG_{IJKL}.}
Notice that in accord with the argument presented at the end of the previous 
subsection, it is indeed the modified field strength $\tilde G_{IJKL}$ 
-- rather than the original $G_{IJKL}$ -- that is better behaved near the 
boundary in the presence of the gluino condensate.  In particular, 
when a gluino condensate $\left\langle\bar\chi^a\Gamma_{ABC}\chi^a
\right\rangle$ forms at the boundary, $\tilde G_{IJKL}$ stays finite and 
continuous in the vicinity of the boundary, while the original field strength 
$G_{IJKL}$ develops a rather singular, compensating vacuum expectation value 
supported at the boundary, $G_{ABC\,11}\sim\delta(x^{11})\left\langle\bar
\chi^a\Gamma_{ABC}\chi^a\right\rangle$.  

The next step is to look at the unbroken supersymmetry conditions, 
$\delta\psi_A=\delta\psi_{11}=\delta\chi^a=0$, with the supersymmetry 
variations $\delta\psi_A$, $\delta\psi_{11}$ and $\delta\chi^a$ given by 
\eemthsusy .  The gluino condensate is of order one at the 
scale where the strong coupling develops in the Yang-Mills sector, therefore 
the contribution of the gluino condensate  to the supersymmetry variations 
\eemthsusy\ is of order $\kappa^{2/3}$.  On the other hand, 
$\tilde G_{IJKL}$ contributes 
already at order $\kappa^0$, but since it only acquires non-zero values of 
order $\kappa^{2/3}$, both effects are of the same order in the 
long-wavelength expansion in the powers of $\kappa^{2/3}$.  

In terms of the redefined fields, the supersymmetry variations \eemthsusy\ 
take the following interesting form:
\foot{Since there are no corrections at this order in $\kappa^{2/3}$ to 
$\delta\chi^a$ \hweff , 
the corresponding equation $\delta\chi^a=0$ is solved at order $\kappa^0$ 
just as in the Calabi-Yau compactifications at weak coupling, and we will 
drop it from now on.}
\eqn\eemthsusynew{\eqalign{\delta\psi_A&=D_A\eta+
\frac{\sqrt 2}{288}\tilde G_{IJKL}\left(\Gamma_A{}^{IJKL}-8\delta_A^I
\Gamma^{JKL}\right)\eta+\ldots,\cr
\delta\psi_{11}&=D_{11}\eta+\frac{\sqrt 2}{288}
\tilde G_{IJKL}\left(\Gamma_{11}{}^{IJKL}-8\delta_{11}^I
\Gamma^{JKL}\right)\eta\cr
&\qquad{}+\frac{1}{192\pi}\left(\frac{\kappa}{4\pi}\right)^{2/3}\delta(x^{11})
\left(\bar\chi^a\Gamma_{ABC}\chi^a\right)\Gamma^{ABC}\eta+\ldots.\cr}}
Here $\ldots$ again denotes terms of order $\kappa^{4/3}$.  

Two aspects of these formulas are worth pointing out:

(1)  The gluino condensate drops out from the supersymmetry variation of 
$\psi_A$, 
and it is therefore the normal component $\psi_{11}$ of the eleven-dimensional 
gravitino that plays the role of the would-be goldstino in the theory.  This 
indicates that the variables of $M$-theory are perhaps better suited for the 
description of the super-Higgs effect in the heterotic string than those of 
the weakly coupled theory.  

(2)  In the supersymmetry variation of $\psi_{11}$, the term bilinear in the 
gluinos is accompanied by a term that depends on the normal derivative of the 
spinor, $D_{11}\eta$.  

These two facts represent yet another ``conspiracy'' in the microscopic 
Lagrangian of $M$-theory on the manifold with boundary, and will be 
crucial in our subsequent analysis of supersymmetry breaking in the presence 
of the gluino condenstate.  In particular, this ``conspiracy'' will allow us 
to solve the unbroken supersymmetry conditions in the vicinity of the boundary 
where the gluino condensate forms.  Indeed, with the gluino condensate 
appearing only in the condition for the vanishing of $\delta\psi_{11}$, 
where the $D_{11}\eta$ term appears, one can now hope to solve these
conditions by allowing $\eta$ to depend on 
$x^{11}$ appropriately.  This is to be contrasted with the analogous situation 
in the theory dimensionally reduced to ten dimensions, which corresponds to 
the weakly coupled heterotic theory.  In the dimensionally reduced theory, 
the $D_{11}\eta$ 
term will be absent from $\delta\psi_{11}$, and supersymmetry will necessarily 
be broken by the gluino condensate in this approximation.  

\subsec{Local Solution of the Unbroken Supersymmetry Conditions}

In the previous section we have noticed that both the equations of motion and 
the Bianchi identity of the modified field strength $\tilde G_{IJKL}$ 
coincide 
with the equations for the unmodified $G_{IJKL}$ in the absence of the 
condensate, and can therefore be solved using the results of \wstr .  
Since $\delta\psi_A$ was also shown to be 
independent of the gluino condensate, we can extend this argument and start 
with any solution of the system of equations studied in \wstr , and 
use it directly to solve our equations {\it in the presence\/} of the 
condensate.  

A solution of the unbroken supersymmetry conditions in the absence of 
the gluino condensate to order $\kappa^{2/3}$ is represented \wstr\ by a 
four-form field strength 
$\CG_{IJKL}$ or order $\kappa^{2/3}$ (which we set equal to our modified 
field strength $\tilde G_{IJKL}$), a metric on $\X\times\S^1/\Z_2$ (which 
differs by effects of order $\kappa^{2/3}$ from the product of the Ricci-flat 
metric on $\X$ and the canonical metric on $\S^1/\Z_2$), and a spinor 
$\tilde\eta$ (which differs from the covariantly constant spinor $\eta_0$ on 
$\X$ by terms of order $\kappa^{2/3}$).  The existence of such a solution in 
the absence of the gluino condensate has been shown in \wstr .  

The formation of a gluino condensate is also an effect of order 
$\kappa^{2/3}$, and will further modify $\tilde\eta$.  On the other hand, 
since the gluino condensate decouples in our modified variables from all 
equations except $\delta\psi_{11}=0$, the four-form 
field strength and the metric will not be modified by the presence of the 
condensate.  

To find a solution of the unbroken supersymmetry conditions in the presence 
of the gluino condensate to order $\kappa^{2/3}$, the last equation that 
remains to be satisfied is $\delta\psi_{11}=0$, or more explicitly
\eqn\eelasteq{\eqalign{D_{11}\eta'&=-\frac{\sqrt 2}{288}\tilde G_{IJKL}
\left(\Gamma_{11}{}^{IJKL}-8\delta_{11}^I\Gamma^{JKL}\right)\eta'\cr
&\qquad{}-\frac{1}{192\pi}\left(\frac{\kappa}{4\pi}
\right)^{2/3}\delta(x^{11})\left(\bar\chi^a\Gamma_{ABC}
\chi^a\right)\Gamma^{ABC}\eta'+\ldots,}}
with $\ldots$ again denoting higher order terms in $\kappa^{2/3}$.  

Given that $\tilde\eta$ solves the equation \eelasteq\ in the absence of the 
gluino condensate, the equation to be actually solved at order $\kappa^{2/3}$ 
is 
\eqn\eecohoeq{\p_{11}(\eta'-\tilde\eta)=-\frac{1}{192\pi}\left(
\frac{\kappa}{4\pi}\right)^{2/3}\delta(x^{11})\left(\bar\chi^a\Gamma_{ABC}
\chi^a\right)\Gamma^{ABC}\eta_0.}
This equation has a very simple solution, 
\eqn\eeunbroke{\eta'=\tilde\eta-\frac{1}{384\pi}\left(\frac{\kappa}{4\pi}
\right)^{2/3}\epsilon(x^{11})\,\left(\bar\chi^a\Gamma_{ABC}\chi^a\right)
\Gamma^{ABC}\eta_0.}
This spinor $\eta'$ -- which differs from $\tilde\eta$ and therefore from the 
covariantly constant spinor $\eta_0$ on $\X$ by terms of order $\kappa^{2/3}$ 
-- thus satisfies the last 
of the unbroken supersymmetry conditions, \eelasteq , in the vicinity of 
the gluino condensate to the required order in $\kappa^{2/3}$.  

Of course, we have to check that $\eta'$ still satisfies the rest of the 
unbroken supersymmetry conditions, $\delta\psi_A=0$.  This is indeed the case 
to order $\kappa^{2/3}$, as the gluino condensate is 
covariantly constant.  Also, for this spinor to be well-defined on the 
eleven-dimensional 
orbifold, it has to be even under the $\Z_2$ action that defines the orbifold, 
\eqn\eechiral{\eta'(-x^{11})=\Gamma_{11}\eta'(x^{11}).}
The $\eta'$ of \eeunbroke\ indeed satisfies this chirality condition, in 
an interesting way.  While $\tilde\eta$ is chiral in ten dimensions and 
satisfies $\Gamma_{11}\tilde\eta=\tilde\eta$, the second term in \eeunbroke\ 
is proportional to $\Gamma_{ABC}\tilde\eta$ which is anti-chiral in ten 
dimensions, $\Gamma_{11}\Gamma_{ABC}\tilde\eta=-\Gamma_{ABC}\tilde\eta$.%
\foot{Thus, the spinor that represents the unbroken supersymmetry does not 
have a definite ten-dimensional chirality; however, it still satisfies the 
chirality condition in four dimensions, $\gamma_5\eta'=\eta'$.}
In $\eta'$, this anti-chiral spinor is however multiplied by the step 
function $\epsilon(x^{11})$ which is odd under the change of orientation of 
the eleventh dimension $x^{11}\rightarrow -x^{11}$. Thus, $\eta'$ is even 
under the combined action of ten-dimensional chirality and orientation 
reversal of the eleventh dimension, and satisfies the orbifold condition 
\eechiral .  Hence, surprisingly enough, the presence 
of the eleventh dimension of $M$-theory has allowed us to solve the unbroken 
supersymmetry conditions in the vicinity of the space-time boundary that 
supports the gluino condensate!  

So far, we haven't taken into account the global topology of the orbifold 
dimension.  Strictly speaking, our analysis therefore shows  that in the 
presence of the gluino condensate, supersymmetry is unbroken in the formal 
limit of 
infinitely strong heterotic string coupling, i.e.\ as we send $R_{11}$ to 
infinity. In this limit, $\eta'$ of \eeunbroke\ would be a  
globally well-defined solution of the unbroken supersymmetry conditions, to 
order $\kappa^{2/3}$.  

\subsec{Global Obstructions and Supersymmetry Breaking}

So far we have seen that even the observer located directly at the boundary 
where the gluino condensate forms will see unbroken supersymmetry, 
as long as the other, weakly coupled boundary is far away.  
Now we will try to extend the local solution \eeunbroke\ of the unbroken 
supersymmetry conditions to a global solution defined everywhere in $\R^4
\times\S^1/\Z_2\times\X$, for finite radius of the orbifold dimension. 

When we try to do so, we encounter an obstruction. 
We have already solved the unbroken supersymmetry conditions at the end with 
the strongly coupled $E_8$ sector, where the gluino condensate forms.  The 
unbroken supersymmetry conditions are also satisfied everywhere in the bulk, 
so they only remain to be satisfied at the weakly coupled $E_6$ end.  
Since there is no gluino condensate at this weakly coupled end, the unbroken 
supersymmetry conditions simply require $\eta'$ to be continuous across 
this boundary, 
\eqn\eeunival{\eta'(-R_{11})=\eta'(R_{11}).}
However, the chirality properties of 
$\eta'$ discussed in the previous subsection can be used to show that the 
condition \eeunival\ 
is violated if the gluino condensate at the strongly coupled end is 
non-zero.  Indeed, while $\tilde\eta$ is even under $x^{11}\rightarrow - 
x^{11}$, the term proportional to $\epsilon(x^{11})$ in $\eta'$ is odd under 
this transformation.  Therefore, a topological obstruction must exist that 
breaks supersymmetry globally, even though we can solve the unbroken 
supersymmetry conditions locally in any chosen coordinate system.  

Now we would like to understand more precisely the nature of this topological 
obstruction.  To do so, it is natural to consider a slightly more general 
case, in which gluino condenstates are allowed to form at both boundaries of 
the space-time manifold.  

Notice first that the gluino condensate $\left\langle\bar
\chi^a\Gamma_{ABC}\chi^a\right\rangle$ is proportional to the components of a 
three-form on $\X$, but it is actually 
better to think of it as a four-form on $\X\times\S^1/\Z_2$.  Indeed, the 
delta function localized at the space-time boundary transforms as the 
$dx^{11}$ component of a one-form whose other components are identically 
zero.  We will write the gluino condensate at the $\alpha$-th component of
 the space-time boundary as a four-form $\omega_{(\alpha)}$, 
in the following coordinate-free way: 
\eqn\eecondform{\omega_{(\alpha)}\equiv\delta(x^{11})dx^{11}\,
\wedge\,\left\langle\bar\chi^a\Gamma_{ABC}\chi^a\right\rangle 
dx^A\wedge dx^B\wedge dx^C.}
Here $\delta(x^{11})$ is the delta function supported at the 
$\alpha$-th boundary component, and $\chi$'s are the corresponding gluinos.  
{}For a covariantly constant condensate, $\omega_{(\alpha)}$  is closed and
$\Z_2$-invariant, and therefore defines a $\Z_2$-equivariant 
cohomology class on $\X\times\S^1/\Z_2$.  More importantly for our purposes, 
$\omega_{(\alpha)}$ is closed under the nilpotent operator $d_{11}\equiv
dx^{11}\p_{11}$ that represents the exterior derivative along the eleventh 
dimension, and we denote by $[\omega_{(\alpha)}]$ the corresponding 
$\Z_2$-equivariant cohomology class in the cohomology defined by 
$d_{11}$.  

{}For \eelasteq\ to have a global solution, the right hand side of equation 
\eecohoeq\ has to be exact with respect to $d_{11}$, as a $\Z_2$-equivariant 
form on $\X\times\S^1/\Z_2$.  Thus, the topological condition that 
allows us to extend the local solution of the unbroken supersymmetry 
conditions to a global one, is that the $\Z_2$-equivariant $d_{11}$-cohomology 
class of $\omega_{(\alpha)}$, summed over all boundary components, vanish: 
\eqn\eecohomcond{\sum_\alpha [\omega_{(\alpha)}]=0.}
In general, this condition is violated, and supersymmetry is broken by the 
global topology of the extra dimension of $M$-theory.  

There is however one simple way to satisfy the cohomological condition 
\eecohomcond , which leads to an {\it a posteriori\/} 
plausibility argument indicating that we could have perhaps 
expected locally ubroken supersymmetry in the presence of a gluino 
condensate in $M$-theory, with supersymmetry broken only by global 
topological effects.%
\foot{This argument was pointed out to me by E. Witten.} 
Set $\tilde G_{IJKL}$ to zero, and consider the case when strong coupling of 
equal strengths develops in the two gauge groups at the two ends of the 
world.  
Now if we put equally strong and opposite gluino condensates at the two 
boudaries, the topological obstruction \eecohomcond\ vanishes, and 
supersymmetry is unbroken.  This can be understood if we start from the limit 
of weak heterotic coupling, described by ten-dimensional heterotic 
supergravity.  Indeed, it is clear that in the presence of two equally strong 
gluino condensates that differ only by a minus sign, supersymmetry stays 
unbroken in ten dimensions.  Next we enlarge the 
string coupling and 
go back to the eleven-dimensional description.  Assuming 
that the mechanism in which supersymmetry is preserved 
is local in the eleventh dimension, 
supersymmetry should be locally preserved in the vicinity of each  
gluino condensate.  Now if we change the value of one of the condensates, 
supersymmetry will be broken, but since it is preserved locally in the 
neighborhood of each condensate, it can only be broken by effects that involve 
the global topology of the orbifold dimension.  
Indeed, it is easy to find the globally 
defined spinor $\eta''$ that represents the unbroken supersymmetry in the 
background of such equally strong but opposite condensates: 
\eqn\eespinsusy{\eta''=\eta_0-\frac{1}{384\pi}\left(\frac{\kappa}{4\pi}
\right)^{2/3}\epsilon(x^{11})\,\left(\bar\chi^a\Gamma_{ABC}\chi^a\right)
\Gamma^{ABC}\eta_0,}
with $\eta_0$ the covariantly constant spinor on $\X$.  Clearly, $\eta''$ 
has a jump at both ends of the world, with opposite values corresponding to 
the strengths of the two gluino condensates.  

In the phenomenologically most interesting compactifications -- notably, those 
with the spin connection embedding that breaks one of the $E_8$'s to $E_6$ 
-- one end of the world supports the grand-unified degrees of freedom that are 
weakly coupled, while the hidden $E_8$ sector is strongly coupled and 
should develop a non-zero gluino condensate.  With a gluino condensate at only 
one end of the world, the cohomology condition \eecohomcond\ cannot be 
satisfied, and supersymmetry is broken by the global topology of the extra 
dimension of $M$-theory, in a mechanism that is reminiscent of the Casimir 
effect.  

\bigskip

I wish to thank Tom Banks, David Gross, Shamit Kachru, Sanjaye Ramgoolam, 
Eric Sharpe, Eva Silverstein, and especially Edward Witten for useful 
discussions and comments.  
Results of this paper have been presented in an invited talk at Strings '96, 
Santa Barbara, July 16, 1996; I would like to thank Joe Polchinski, Shyamoli 
Chaudhuri, Gary Horowitz and L\'arus Thorlacius for their excellent 
organization of the conference.  
The paper was completed at Aspen Center for Physics, whose kind hospitality 
is gratefully acknowledged.  
This work was supported in part by NSF Grant PHY90-21984.  
\listrefs
\end